\documentclass[12pt]{article}
\usepackage{amsmath, amssymb}
\usepackage{geometry}
\usepackage{setspace}
\usepackage{hyperref}
\usepackage{graphicx}
\usepackage{subcaption}

\begin{document}

\title{Gamma Distribution for Equilibrium Analysis of Discrete Stochastic Logistic Population Models}
\author{Haiyan Wang\\
School of Mathematical and Natural Sciences\\
Arizona State University\\
Phoenix, AZ 85069\\
haiyan.wang@asu.edu
}
\date{}
\maketitle

\begin{abstract}
Stochastic models play an essential role in accounting for the variability and unpredictability seen in real-world. This paper focuses on the application of the gamma distribution to analysis of the stationary distributions of populations governed by the discrete stochastic logistic equation at equilibrium.  It is well known that the population dynamics of deterministic logistic models are dependent on the range of intrinsic growth rate. In this paper, we identify the same feasible range of the intrinsic growth rate for the stochastic model at equilibrium and establish explicit mathematical relation among the parameters of the gamma distribution and the stochastic models. We analyze the biological implications of these relationships, with particular emphasis on how the shape and scale parameters of the gamma distribution reflect population dynamics at equilibrium. These mathematical relations describe the impact of the variance of the stochastic perturbation on the intrinsic growth rate, and, in particular, reveal that there are two branches of the intrinsic growth rates representing alternative stable states at equilibrium.   
\end{abstract}

\section{Introduction}
Modeling population dynamics is a cornerstone of ecology and biology, as it allows researchers to understand how populations grow, stabilize, or decline under various environmental pressures. The logistic difference equation has been widely used to model population growth under resource limitations. In its deterministic form, the logistic equation captures how populations initially grow exponentially, then slow down as they approach the carrying capacity, which reflects the maximum population size that the environment can support.  The discrete deterministic logistic difference equation after scaling  is expressed as:

\begin{equation} \label{eq:logistic1}
x_{t+1} = r x_t \left(1 - x_t \right)
\end{equation}
where \( x_t \) represents the population size at time \( t \), \( r \) is the intrinsic growth rate.  This equation describes the population's approach to a stable equilibrium under idealized conditions.

However, real-world populations are rarely deterministic; they experience random fluctuations due to environmental variability, demographic stochasticity, and other unpredictable factors. These stochastic effects motivate the need for stochastic versions of the logistic equation.  To address these real-world complexities, the stochastic logistic equation introduces randomness into the population dynamics. One considers the stochastic version of (\ref{eq:logistic1})

\begin{equation} \label{eq:logistic2}
X_{t+1} = r X_t \left( 1 - X_t \right) \epsilon_t
\end{equation}
where \( X_t \) is a distribution of population size at a specific time \( t \), $\epsilon_t$ is a small nonnegative perturbation distribution representing some stochastic effects, assumed its mean is $1$ ($E[\epsilon_t] = 1$) and independent of $X_t$. Stochastic models offer significant advantages in capturing the dynamics of real-world populations. Unlike deterministic models, which predict fixed outcomes, stochastic models incorporate the inherent randomness and variability observed in natural systems. Stochastic models allow for the prediction of distributions of population sizes rather than precise point estimates, offering insights into both the average behavior and the range of possible fluctuations around it \cite{Allen, Renshaw, Matis2003,May2001,Dennis2016,Nasell2003,kot2001elements,Kemp1993}. There are a rich literature on the study of the stochastic effects for various discrete models \cite{Braverman2013,Aktar2023,Schreiber2021}.  

In this paper, we focus on analysis of stochastic logistic difference models at equilibrium with the gamma distribution. In contrast to deterministic models, in which typically predict point equilibria, stochastic population models often predict stationary distributions around a steady state. The gamma distribution, in particular, has been shown to provide an accurate approximation of these stationary distributions in various ecological contexts \cite{Dennis1984,Pielou1975,Engen1978}. The successful application of the gamma distribution is especially evident in studies of species such as the Tribolium beetle, where populations in laboratory settings tend to fluctuate around a mean equilibrium size due to environmental variability \cite{Dennis1984,Peters1989,DennisCostantino1988,Costantino1981}. 

On the other hand, it is well known that the population dynamics of the deterministic logistic model (\ref{eq:logistic1}) such as stable growth, periodic oscillations, chaos, or extinction, are dependent on the value of \( r \). In particular, for \( 1 < r < 3 \), the population governed by \eqref{eq:logistic1} grows and eventually reaches a nonzero steady
state \cite{may1976simple,strogatz2018nonlinear,hilborn2000chaos}. In this paper we identify the same feasible range of $r$ ($1< r < 3$) for the stochastic model (\ref{eq:logistic2}) at equilibrium,  establish explicit mathematical relation of $r$ with the parameters of the gamma distribution,  and analysis the impact of the variance of $\epsilon_t$ on equilibrium states. In particular, the mathematical formulation suggests that there are two branches of the intrinsic growth rate, \( r_+ \) and \( r_- \), represent alternative stable states, corresponding to higher a growth rate and lower growth rate.  In addition, the mathematical relation of \( r \) is independent from \( \theta \), the scale parameter of the gamma distribution, suggesting that the intrinsic growth rate is primarily a function of the population's internal dynamics (reflected by \( k \)) rather than the scale of the population size distribution (governed by \( \theta \)). This emphasizes the role of internal biological mechanisms in driving growth, regardless of the absolute population size or its spread.

Our key contributions in this paper include: 
1. Identifying the explicit range of the intrinsic growth rate $r$ for the stochastic models at equilibrium, consistent with the classical results on the deterministic logistic equation. 2. Establishing the mathematical relation of $r$ with the parameters of the gamma distribution, investigating the impact of the perturbation $\epsilon_t$ on $r$,  providing theoretical insights into population dynamics of the stochastic models compared to the determinist logistic model. 3. Identify two branches of the intrinsic growth rate, \( r_+ \) and \( r_- \) for (\ref{eq:logistic2}), representing alternative stable states and corresponding to higher a growth rate and lower growth rate. 4. Analyzing the biological implications of these relationships, with particular emphasis on how the shape and scale parameters of the gamma distribution impact population dynamics at equilibrium. 

As a result, we provide a framework for understanding the role of the gamma distribution in modeling stationary distributions for populations fluctuating around a stable equilibrium.

\section{Equilibrium Analysis of Discrete Logistic Equation}

\subsection{The Gamma Distribution at Equilibrium}

The gamma distribution is defined on the interval $(0, \infty)$ and is characterized by two positive parameters: the shape parameter $k$ and the scale parameter $\theta$. Its probability density function (PDF) is:

\begin{equation} \label{eq:gamma_pdf}
f(x; k, \theta) = \frac{x^{k - 1} e^{-x / \theta}}{\Gamma(k) \theta^k}, \quad \text{for } x > 0,
\end{equation}
where $\Gamma(\cdot)$ is the gamma function. For positive integer $n$, $\Gamma(n)=(n-1)!$.  In the context of the gamma distribution, the parameter $k$ (sometimes referred to as the shape parameter) determines the shape and skewness of the distribution, while the scale parameter $\theta$ controls the spread of the distribution. This distribution is used in many areas, such as population dynamics, waiting times for events, and biological processes \cite{wikipedia_gamma}. 

From a theoretical perspective, the gamma distribution emerges as a stationary solution to various stochastic processes, such as birth-death processes and stochastic differential equations (SDEs), that incorporate randomness into population dynamics. These stochastic models introduce random fluctuations around a deterministic growth process, allowing for a more realistic representation of natural populations. At the equilibrium of stochastic logistic equations, the statistical properties of the population become stable over time. The gamma distribution's ability to describe populations that stabilize around an equilibrium point under stochastic effects makes it a compelling tool for modeling real-world population dynamics. The gamma distribution has been shown to provide an accurate approximation of these stationary distributions in various ecological contexts \cite{Dennis1984,Peters1989,DennisCostantino1988,Costantino1981}. We assume that the population size $X_t$ in \eqref{eq:logistic2} at equilibrium follows a gamma distribution with parameters $k$ and $\theta$:
\begin{equation} \label{eq:x_gamma_distribution}
X_t \sim \text{Gamma}(k, \theta).
\end{equation}
The mean and variance of the gamma distribution are:
\begin{align}
\mu &= E[X_t] = k \theta, \label{eq:gamma_mean} \\
\sigma^2 &= \text{Var}[X_t] = k \theta^2. \label{eq:gamma_variance}
\end{align}
The first two moments are:
\begin{align}
E[X_t] &= \mu = k \theta, \label{eq:mean_mu} \\
E[X_t^2] &= k (k + 1) \theta^2. \label{eq:second_moment}
\end{align}
The higher order moments of \( X_t \) are:
\begin{align}
E[X_t^n] &= \theta^n \frac{\Gamma(k+n )}{\Gamma(k)}, \label{eq:gamma_moment} \\
\end{align}

Because of the nonlinearity in (\ref{eq:logistic2}), $X_{t+1}$ may not necessarily follow the gamma distribution. However, it is reasonable to assume that the population at equilibrium from specific time $t$ to $t+1$ maintains the same expectation and variance. Therefore, at equilibrium, we assume that, for a specific time $t$,  
\begin{equation} \label{eq:conditions}
E[X_{t+1}] = E[X_t], \,\,\,  Var[X_{t+1}] = Var[X_t] 
\end{equation}
which allows us to derive explicit mathematical relations for $r$ in terms of $k$ and $\theta$. These mathematical relations further confirm the gamma distribution’s flexibility in accommodating skewed distributions and its ability to model variability around an equilibrium make it particularly suitable for representing the outcomes of stochastic processes in ecology.

\subsection{Expectation Condition at Equilibrium}
Since $\epsilon_t$ is independent, we have  
\begin{equation} \label{eq:expected_value}
E[X_{t+1}] = E\left[ r X_t \left(1 - X_t \right )  \epsilon_t \right]=E\left[ r X_t \left(1 - X_t \right ) \right] E[\epsilon_t].
\end{equation}
Since $E[\epsilon_t]=1$, and at equilibrium, ($E[X_{t+1}] = E[X_t] = \mu$), we have
\begin{equation} \label{eq:expanded_equation}
\mu = r \left( \mu - E[X_t^2] \right).
\end{equation}
Subtract $r \mu$ from both sides and simplify it:
\begin{equation} \label{eq:simplified_equation}
\mu (1 - r) = - r E[X_t^2].
\end{equation}
Substituting $\mu = k \theta$ and $E[X_t^2]=\sigma^2 + \mu^2$ into the equilibrium equation \eqref{eq:simplified_equation}:
\begin{equation} \label{eq:equilibrium_in_k_theta}
k \theta (1 - r) = - r \left( k \theta^2 + k^2 \theta^2 \right).
\end{equation}
Divide both sides by $k \theta$ (since $k, \theta > 0$):

\begin{equation} \label{eq:equilibrium_divided}
(1 - r) = - r \theta (1 + k).
\end{equation}
Solve for $\theta$:

\begin{equation} \label{eq:theta_solution}
\theta = \frac{r - 1}{r (1 + k)}.
\end{equation}
Therefore, this suggests that for $r > 1$, this equilibrium may exist under the gamma distribution assumption.

\subsection{Variance Condition at Equilibrium}
In addition to the mean, we can compare the variance condition at equilibrium $$ \text{Var}(X_{t+1}) = \text{Var}(X_t).$$ This provides additional conditions that can help determine the gamma distribution parameters. To compute \( \text{Var}(X_{t+1}) \), we need \( E[X_{t+1}] \) and \( E[X_{t+1}^2] \):

\[
\text{Var}(X_{t+1}) = E[X_{t+1}^2] - \left( E[X_{t+1}] \right)^2.
\]
We already have \( E[X_{t+1}] = \mu = k \theta \). To compute \( E[X_{t+1}^2] \), we expand \( X_{t+1}^2 \):

\begin{align*}
X_{t+1}^2 &= \left( r X_t - r X_t^2  \right)^2 \epsilon_t^2 \\
&= \left (r^2 X_t^2 - 2 r^2 X_t^3 + r^2 X_t^4 \right )\epsilon_t^2
\end{align*}
We take expectations:
\begin{equation} \label{eq:expectation_xt1_squared}
E[X_{t+1}^2] = \left ( r^2 E[X_t^2] - 2 r^2 E[X_t^3 ] + r^2 E[X_t^4] \right )E[\epsilon_t^2].
\end{equation}
At equilibrium, \( \text{Var}(X_{t+1}) = \text{Var}(X_t) = k \theta^2 \). Therefore,

\begin{equation} \label{eq:variance_equilibrium}
E[X_{t+1}^2] - (k \theta)^2 = k \theta^2
\end{equation}
and
\begin{equation} \label{eq:variance_equilibrium_simplified}
E[X_{t+1}^2] = k \theta^2 (1 + k).
\end{equation}
We can cancel \( \theta^2 \) from both sides as \( \theta > 0 \):

\begin{equation} \label{eq:variance_equilibrium_theta_cancelled}
\frac{E[X_{t+1}^2]}{\theta^2} = k (1 + k).
\end{equation}
Compute the required terms:

\begin{align*}
E[X_t^2] &= \theta^2 k (k + 1), \\
E[X_t^{3}] &= \theta^{3} \frac{\Gamma(k + 3)}{\Gamma(k)} = \theta^{3} (k + 2)(k + 1)k, \\
E[X_t^{4}] &= \theta^{4} \frac{\Gamma(k + 4)}{\Gamma(k)} = \theta^{4} (k + 3)(k + 2)(k + 1)k.
\end{align*}
With cancellation of $\theta^2$,  \eqref{eq:variance_equilibrium_theta_cancelled} and \eqref{eq:expectation_xt1_squared} give
\begin{align*}
& \Big( r^2 k (k + 1) - 2 r^2 \theta (k + 2)(k + 1)k + r^2 \theta^{2}  (k + 3)(k + 2)(k + 1)k \Big ) E[\epsilon_t^2] = k (1 + k).
\end{align*}
Dividing by $E[\epsilon_t^2]$ and subtracting $k (k + 1)$ at both sides give  
\begin{align*}
&  (r^2-1) k (k + 1) - 2 r^2 \theta (k + 2)(k + 1)k + r^2 \theta^{2}  (k + 3)(k + 2)(k + 1)k = \frac{k (1 + k)}{E[\epsilon_t^2]}-k(k+1).
\end{align*}
Multiplying by $-1$ at both sides produces 
\begin{align*}
&  (1-r^2) k (k + 1) + 2 r^2 \theta (k + 2)(k + 1)k - r^2 \theta^{2}  (k + 3)(k + 2)(k + 1)k = \frac{E[\epsilon_t^2]-1}{E[\epsilon_t^2]}k(k+1).
\end{align*}
Since $Var[\epsilon_t]=E[\epsilon_t^2]-E[\epsilon_t]^2=E[\epsilon_t^2]-1$, 
\begin{align*}
&  (1-r^2) k (k + 1) + 2 r^2 \theta (k + 2)(k + 1)k - r^2 \theta^{2}  (k + 3)(k + 2)(k + 1)k =\frac{Var[\epsilon_t]}{Var[\epsilon_t]+1}k(k+1).
\end{align*}
%
%
%
%
%
Note that $\theta = \frac{(r - 1)}{r (1 + k)}$, therefore,
\begin{equation}
(1 - r^2) k (k + 1) + 2 r (r - 1) k (k + 2) - (r - 1)^2 k \cdot \dfrac{ (k + 2)(k + 3) }{ k + 1 } = \frac{Var[\epsilon_t]}{Var[\epsilon_t]+1}k(k+1)
\label{eq:simplified_main_n1}
\end{equation}
Dividing both sides by \( k (r - 1) \), we have

\begin{equation}
\dfrac{ (1 - r^2) k (k + 1) }{ k (r - 1) } + \dfrac{ 2 r (r - 1) k (k + 2) }{ k (r - 1) } - \dfrac{ (r - 1)^2 k \cdot \dfrac{ (k + 2)(k + 3) }{ k + 1 } }{ k (r - 1) } = \frac{Var[\epsilon_t]}{Var[\epsilon_t]+1}\frac{k+1}{r - 1}
\label{eq:divided_n1}
\end{equation}
After simplification, equation \eqref{eq:divided_n1} becomes:
\begin{equation}
- (r + 1)(k + 1) + 2 r (k + 2) - \dfrac{ (r - 1) (k + 2)(k + 3) }{ k + 1 } = \frac{Var[\epsilon_t]}{Var[\epsilon_t]+1}\frac{k+1}{r - 1}
\label{eq:divided_simplified_n1}
\end{equation}
Multiply equation \eqref{eq:divided_simplified_n1} by \( k + 1 \):
\begin{equation}
- (r + 1)(k + 1)^2 + 2 r (k + 2)(k + 1) - (r - 1)(k + 2)(k + 3) = \frac{Var[\epsilon_t]}{Var[\epsilon_t]+1}\frac{(k+1)^2}{r - 1}
\label{eq:multiplied_n12}
\end{equation}
The left side of \eqref{eq:multiplied_n12} is 
$$
- r k^2 - 2 r k - r - k^2 - 2k - 1 + 2 r k^2 + 6 r k + 4 r - r k^2 - 5 r k - 6 r + k^2 + 5k + 6 
$$
Thus, with simplification of the left side of \eqref{eq:multiplied_n12}, we arrive at 
\begin{equation} 
-r(k+3)+3k+5 =  \frac{Var[\epsilon_t]}{Var[\epsilon_t]+1}\frac{(k+1)^2}{r - 1}
\label{eq:multiplied_nr}
\end{equation}
Since $r>1$, it follows that 
\begin{equation}
-r(k+3)+3k+5 \ge 0 
\label{eq:multiplied_n1}
\end{equation}
and 
\begin{equation}
1<r \leq \frac{3k+5}{k+3}<3 
\label{eq:multiplied_q}
\end{equation}
Multiply both sides of \eqref{eq:multiplied_nr} by \( (r - 1) \):
\begin{equation}
(r - 1)\Big( -r(k + 3) + 3k + 5 \Big) =  \frac{Var[\epsilon_t]}{Var[\epsilon_t]+1}(k+1)^2
\label{eq:multiplied_qq}
\end{equation}
Simplification of the left side of \eqref{eq:multiplied_qq} leads to:
\begin{equation}
-r^2(k + 3) + 4kr + 8r - 3k - 5 =  \frac{Var[\epsilon_t]}{Var[\epsilon_t]+1}(k+1)^2
\label{eq:multiplied_qq1}
\end{equation}
Now express \eqref{eq:multiplied_qq1} as a quadratic equation in terms of \( r \):
\begin{equation}
(k + 3) r^2 - \left( 4k + 8 \right) r + \left( 3k + 5 + \dfrac{Var[\epsilon_t]}{Var[\epsilon_t] + 1}(k + 1)^2 \right) = 0.
\label{eq:multiplied_qq12}
\end{equation}
Let \( A \), \( B \), and \( C \) in \eqref{eq:multiplied_qq12} for computing its discriminant as:
\begin{align*}
A = k + 3, B = -\left( 4k + 8 \right) = -4k - 8, C = 3k + 5 + \dfrac{Var[\epsilon_t]}{Var[\epsilon_t] + 1}(k + 1)^2
\end{align*}
and
\begin{align*}
B^2 &= 16k^2 + 64k + 64,\\
4AC &= 4(k + 3)\left( 3k + 5 + \dfrac{Var[\epsilon_t]}{Var[\epsilon_t] + 1}(k + 1)^2 \right) \\
&= 4(k + 3)(3k + 5) + \dfrac{4Var[\epsilon_t](k + 3)(k + 1)^2}{Var[\epsilon_t] + 1}\\
  &=12k^2 + 56k + 60 + \dfrac{4Var[\epsilon_t](k + 3)(k + 1)^2}{Var[\epsilon_t] + 1}.
\end{align*}
Thus the discriminant \( D=B^2-4AC \) of the quadratic equation with respect to $r$ is given by:
\begin{align*}
D &= \left( 16k^2 + 64k + 64 \right) - \left( 12k^2 + 56k + 60 + \dfrac{4Var[\epsilon_t](k + 3)(k + 1)^2}{Var[\epsilon_t] + 1} \right) \\
&= 4k^2 + 8k + 4 - \dfrac{4Var[\epsilon_t](k + 3)(k + 1)^2}{Var[\epsilon_t] + 1}\\
&=  4(k + 1)^2 \left( \dfrac{1 - Var[\epsilon_t](k + 2)}{Var[\epsilon_t] + 1} \right).
\end{align*}
For real solutions, the discriminant \( D \) must be non-negative. Therefore, the condition for real solutions is:
\begin{equation}
0 < k \leq \dfrac{1}{Var[\epsilon_t]}-2.
\label{eq:multiplied_q1}
\end{equation}
We now have the maximum of the feasible variance of $\epsilon_t$ 
\begin{equation}
0 \leq Var[\epsilon_t] \leq 0.5.
\label{eq:multiplied_q2}
\end{equation}
Using the quadratic formula, we have
\begin{align}
r_{\pm} &= \dfrac{ 4k + 8 \pm 2(k + 1) \sqrt{ \dfrac{1 - Var[\epsilon_t](k + 2)}{Var[\epsilon_t] + 1} } }{ 2(k + 3) } \\
&= \dfrac{ 2k + 4 \pm (k + 1) \sqrt{ \dfrac{1 - Var[\epsilon_t](k + 2)}{Var[\epsilon_t] + 1} } }{ k + 3 }.
\label{eq:multiplied_q3}
\end{align}
It is easy to see that both $r_{\pm} \geq 1$.   Since $r_{+} > r_{-}$ and $\dfrac{1 - Var[\epsilon_t](k + 2)}{Var[\epsilon_t] + 1} \leq 1$ and therefore 
$$2k + 4 - (k + 1) \sqrt{ \dfrac{1 - Var[\epsilon_t](k + 2)}{Var[\epsilon_t] + 1} } \geq  k + 3 $$

\subsection{Biological Interpretation}
\subsubsection{$Var[\epsilon_t]=0$}
\begin{figure}[h!]
    \centering
    \includegraphics[width=0.6\textwidth]{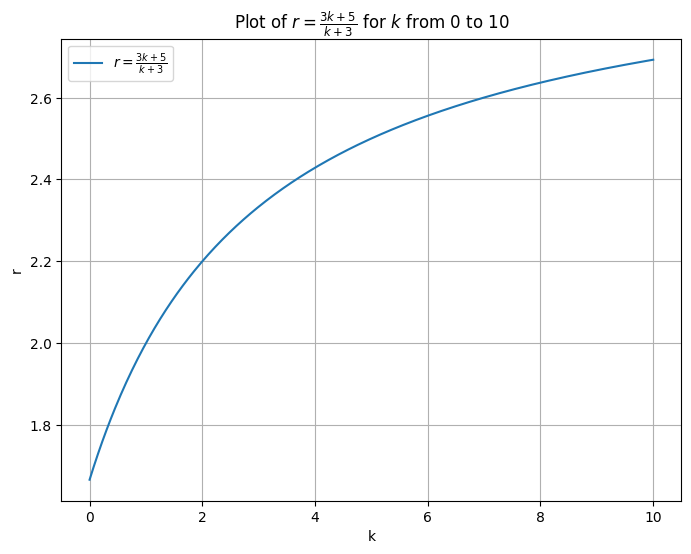}
    \caption{Plot of $r$ in terms of $k$ }
    \label{plot10}
\end{figure}

When the variance of $\epsilon_t$  is zero or sufficient small, equation (\ref{eq:multiplied_nr}) becomes
\begin{equation}
-r(k+3)+3k+5 = 0
\label{eq:multiplied_nr1}
\end{equation}
We can solve for $r$ in terms of $k$
\begin{equation}
r = \frac{ 3k + 5 }{ k + 3 } 
\label{eq:expression_r}
\end{equation}
From equation (\ref{eq:expression_r}), it follows that 
$$
\frac{5}{3}<r<3. 
$$
When $Var[\epsilon_t]$ is sufficient small or zero,  we have the explicit expression of $r$ in terms of $k$, (\ref{eq:expression_r}) and the refined restriction on $r$: $\frac{5}{3} <r<3$. The expression for the growth rate $r$ in terms of $k$ is given by (\ref{eq:expression_r}) and in Figure \ref{plot10}. Upon calculating the derivative of \( r \) with respect to \( k \), we find:
\begin{equation}
\frac{dr}{dk} = \frac{4}{(k + 3)^2}
\end{equation}
This derivative indicates that \( r \) increases with \( k \), but the rate of increase slows down as \( k \) becomes larger. For the case when the relative variance of $\epsilon_t$ is sufficient small, since \( r \) is entirely determined by \( k \), the shape parameter, it indicates that the dynamics of population growth depend on the underlying structure of the population rather than its absolute size. This can be particularly relevant in biological systems where growth is more strongly tied to population interactions (e.g., competition, mating availability) rather than just the magnitude of the population at any given time.

It is well known that the population dynamics of the deterministic logistic model (\ref{eq:logistic1}), such as stable growth, periodic oscillations, chaos, or extinction, are dependent on the value of \( r \). If \( 1 < r < 3 \), the population grows and eventually reaches a nonzero steady state\cite{may1976simple,strogatz2018nonlinear,hilborn2000chaos}.

For the stochastic logistic equation (\ref{eq:logistic2}), the feasible range of $r$, $ (\frac{5}{3}, 3) \subset (1,3) $, at equilibrium implies a form of population regulation, where the population does not experience explosive growth that would destabilize the system. The population may begin to oscillate around the carrying capacity, but these oscillations are damped over time. Eventually, the population settles at the equilibrium. From a biological perspective, this can represent species that have evolved mechanisms for self-regulation, such as density-dependent factors like competition for resources or social structures that limit reproductive success at high population densities. This range of \( r \) suggests that while the population has enough reproductive capacity to grow quickly, it also has internal or external checks that prevent it from overshooting the carrying capacity by a large margin. This could be relevant for species with moderate growth rates that balance reproduction with survival, ensuring long-term persistence in stable environments.

\subsubsection{Impact of $\text{Var}(\epsilon_t)$}
From the discussion in the above section, at equilibrium, the parameters must satisfy  $0 \leq Var[\epsilon_t] \leq 0.5$ and $ 0< k \leq \dfrac{1}{Var[\epsilon_t]}-2$, and the explicit expression of $r$ in terms of $k, \text{Var}(\epsilon_t) $   
\begin{align}
r_{\pm} &= \dfrac{ 2k + 4 \pm (k + 1) \sqrt{ \dfrac{1 - Var[\epsilon_t](k + 2)}{Var[\epsilon_t] + 1} } }{ k + 3 }.
\label{eq:multiplied_q35}
\end{align}

\begin{figure}[h!]
    \centering
    \begin{subfigure}[b]{0.45\textwidth}
        \centering
        \includegraphics[width=\textwidth]{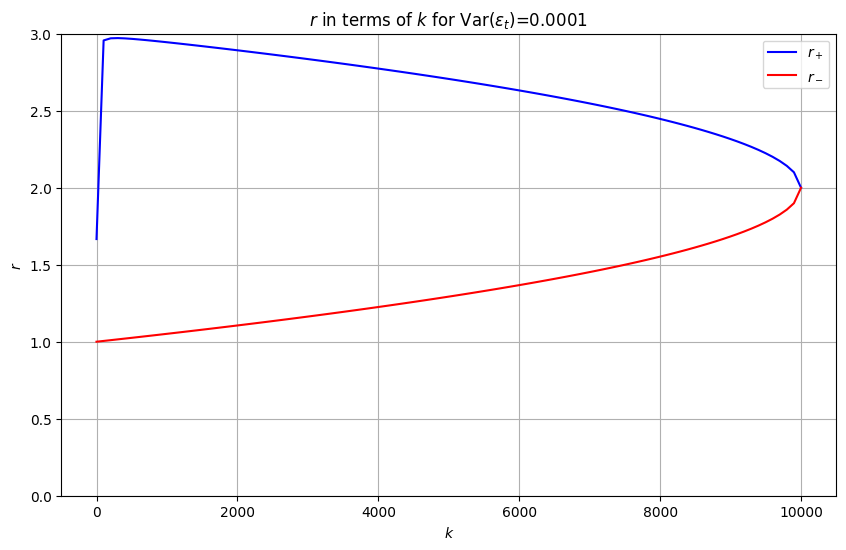}
        \label{fig:sub1}
    \end{subfigure}
    \hfill
    \begin{subfigure}[b]{0.45\textwidth}
        \centering
        \includegraphics[width=\textwidth]{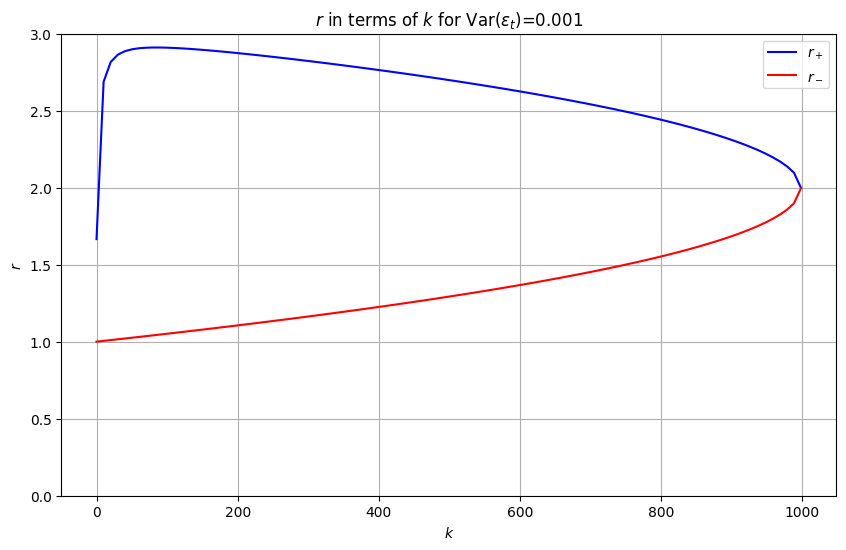}
        \label{fig:sub2}
    \end{subfigure}
    
    \vskip\baselineskip
    
    \begin{subfigure}[b]{0.45\textwidth}
        \centering
        \includegraphics[width=\textwidth]{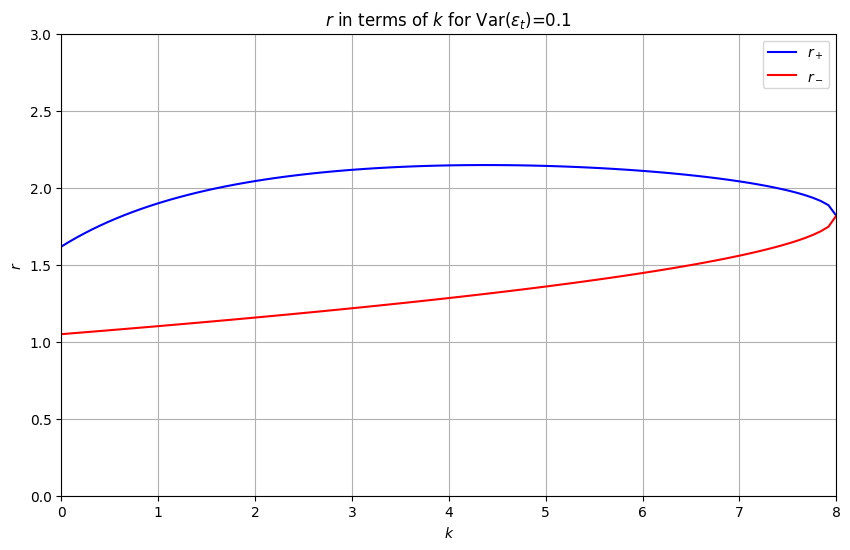}
        \label{fig:sub3}
    \end{subfigure}
    \hfill
    \begin{subfigure}[b]{0.45\textwidth}
        \centering
        \includegraphics[width=\textwidth]{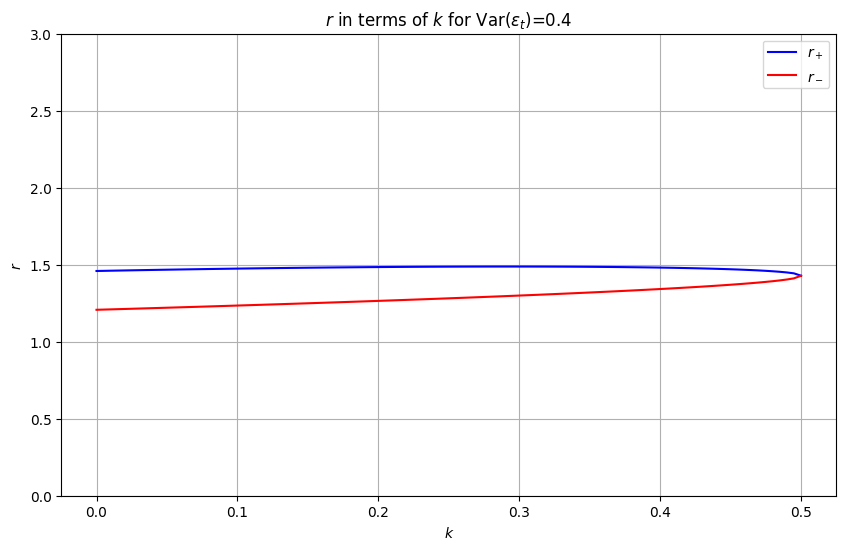}
        \label{fig:sub4}
    \end{subfigure}
    
    \caption{Plots of $r$ in term of $k$}
    \label{fig:main}
\end{figure}

We would like to see its impact on $r$ through simulations in Figure \ref{fig:main}. The four figures illustrate the relationship between the growth rate \( r \) and the parameter \( k \) under different values \( \text{Var}(\epsilon_t) \). 
  
First, we note that all values of \( r \) in Figure \ref{fig:main} are within the range \( (1, 3) \), as expected for the both deterministic and stochastic logistic models. In particular, for the deterministic logistic model, populations grow and stabilize in the same range, where the growth rate balances with the carrying capacity, allowing the population to approach a stable state \cite{may1976simple,strogatz2018nonlinear,hilborn2000chaos}.

\eqref{eq:multiplied_q35} indicates there are two branches, \( r_+ \) and \( r_- \) of the intrinsic growth rate $r$, representing alternative stable states as shown in Figure \ref{fig:main}. The branch \( r_+ \) generally corresponds to higher growth rates and a more resilient population, suggesting a higher-density equilibrium. This state might be stable for species that grow in environments with sufficient resources. In contrast, the branch \( r_- \) represents a lower growth rate, where populations may be more vulnerable to fluctuations. Populations at \( r_- \) might represent a low-density equilibrium, which could be more susceptible to environmental variability and potentially at risk of extinction under high environmental variance.

The condition \( 0 < k \leq \frac{1}{\text{Var}[\epsilon_t]} - 2 \) suggests that the feasible range of \( k \) values is restricted by the environmental variance \( \text{Var}(\epsilon_t) \). Smaller values of \( \text{Var}(\epsilon_t) \) extend the range of \( k \), which implies that populations with larger \( k \) (more symmetrical and regular population distribution) are better supported in stable environments with low variability. As we can see from Figure \ref{fig:main} that for most $k$, in particular when it is not too large, $r$ is increasing with respect to $k$. As a result, smaller values of \( \text{Var}(\epsilon_t) \) indicate more potential for a stable equilibrium across a broader range of \( r \). Thus, the shape parameter \( k \) plays a significant role in defining the equilibrium conditions for populations in fluctuating environments.

For smaller values of \( \text{Var}(\epsilon_t) \), such as \( \text{Var}(\epsilon_t) = 0.0001 \) and \( \text{Var}(\epsilon_t) = 0.001 \), the feasible range of \( k \) is extensive. Populations in this setting are expected to exhibit a wide range of equilibrium growth rates, with both \( r_+ \) and \( r_- \) branches present.  A higher \( k \) value under low variance suggests that the population distribution is less skewed and more resilient, with potential for a stable equilibrium across a broader range of \( r \). Populations with this setup are analogous to species in stable environments that experience minimal environmental variability, allowing them to maintain predictable population dynamics around the equilibrium.

As \( \text{Var}(\epsilon_t) \) increases (e.g., \( \text{Var}(\epsilon_t) = 0.1 \) and \( \text{Var}(\epsilon_t) = 0.4 \)), the range of \( k \) values for stable solutions shrinks. This indicates that populations in fluctuating environments with high variance are limited in terms of the shape parameter \( k \), which restricts their capacity to achieve stable equilibria.

The independence of \( r \) from \( \theta \) suggests that the intrinsic growth rate is primarily a function of the population's internal dynamics (reflected by \( k \)) rather than the scale of the population size distribution (governed by \( \theta \)). This emphasizes the role of internal biological mechanisms in driving growth, regardless of the absolute population size or its spread.

\section{Conclusion and Discussion}

In this work, we explored the use of the gamma distribution to study the populations governed by discrete a stochastic logistic equation at steady state. We investigated mathematical insights into the relation between gamma distribution parameters and the stochastic logistic difference equation at equilibrium, and derived explicit relationships between the parameters of the gamma distribution and the intrinsic growth rate \(r\) of the stochastic logistic difference equation and the variance of small perturbation; and identified a feasible range of $r$ for the stochastic model at equilibrium, which is consistent with the classical results on the dynamics of the deterministic logistic equation. In addition, we provided the ecological interpretations of these relations and their implications from an ecological perspective. For example, we identified there are two branches, \( r_+ \) and \( r_- \) of the intrinsic growth rate $r$ at equilibrium representing alternative stable states.

Several future directions of research can address the challenges identified in this study and extend the current work. Future work will extend the analysis to more complex discrete stochastic models that include additional biological factors, such as age structure, migration, or environmental effects. These models may require the development of new theoretical tools to account for the complexity of real-world population dynamics. A more rigorous mathematical framework is needed to prove the relationships between the parameters of stochastic models and their stationary distributions. These mathematical results will provide deeper insights into the behavior of populations governed by stochastic dynamics and may help identify conditions under which the gamma distribution (or other distributions) can provide valid approximations. 

With appropriate scaling, \eqref{eq:logistic2} can incorporate a parameter for carrying capacity. Alternative distributions that respect the carrying capacity of population sizes, such as the beta distributions, might provide more suitable models for describing stationary distributions in such systems. These alternative approaches could allow for more accurate modeling of real-world population dynamics while retaining key characteristics like variability and skewness.

Finally, future work could use real-world ecological population data to validate the theoretical relations of the stochastic models with the gamma distribution. This validation will involve fitting the model to empirical data, determining the parameters of the stochastic models and the gamma distribution, comparing them with theoretical results in this paper. This process will help ensure that the theoretical results align with the real-world behavior of fluctuating populations.


\begin{thebibliography}{9}

\bibitem{Allen}
Allen, L. J. S. (2010). \emph{An Introduction to Stochastic Processes with Applications to Biology}. Chapman and Hall/CRC.

\bibitem{Renshaw}
Renshaw, E. (1991). \emph{Modelling Biological Populations in Space and Time}. Cambridge University Press.

\bibitem{May2001}
R. M. May, (1974) \emph{Stability and Complexity in Model Ecosystems}, Princeton U.P., Princeton.

\bibitem{Dennis1984}
B. Dennis and G. P. Patil, \emph{The Gamma Distribution and Weighted Multimodal Gamma Distributions as Models of Population Abundance}, Mathematical Biosciences, 68, 187–212, 1984.

\bibitem{Peters1989}
C. S. Peters, M. Mangel, R. F. Costantino, \emph{Stationary Distribution of Population Size in Tribolium}, Bulletin of Mathematical Biology, 51(5), 625–638, 1989.

\bibitem{DennisCostantino1988}
B. Dennis and R. F. Costantino, \emph{Analysis of Steady-State Populations with the Gamma Abundance Model and its Application to Tribolium}, Ecology, 69(4), 1200–1213, 1988.

\bibitem{Costantino1981}
R. F. Costantino and R. A. Desharnais, \emph{Gamma Distributions of Adult Numbers for Tribolium Populations in the Region of their Steady-States}, Journal of Animal Ecology, 50, 667–681, 1981.


\bibitem{Matis2003}
James H. Matis, Thomas R. Kiffe b, Eric Renshawc, Janet Hassan, \emph{A simple saddlepoint approximation for the equilibrium
distribution of the stochastic logistic model of population growth}, Ecological Modelling, 161 (2003) 239–248.


\bibitem{wikipedia_gamma}
Wikipedia contributors, "Gamma distribution," \emph{Wikipedia, The Free Encyclopedia}, 2024.


\bibitem{kot2001elements}
M. Kot, \emph{Elements of Mathematical Ecology}. Cambridge University Press, 2001.

\bibitem{Dennis2016}
Dennis, B., Assas, L., Elaydi, S. et al. \emph{Allee effects and resilience in stochastic populations}. Theor Ecol, 9, 323–335 (2016). https://doi.org/10.1007/s12080-015-0288-2

\bibitem{Kemp1993}
Kemp, William P., and Brian Dennis. “Density Dependence in Rangeland Grasshoppers (Orthoptera: Acrididae).” Oecologia 96, no. 1 (1993): 1–8. http://www.jstor.org/stable/4220493.

\bibitem{may1976simple} May, R. M. (1976). \emph{Simple mathematical models with very complicated dynamics.} \emph{Nature}, 261(5560), 459-467.

\bibitem{strogatz2018nonlinear} Strogatz, S. H. (2018). \emph{Nonlinear Dynamics and Chaos: With Applications to Physics, Biology, Chemistry, and Engineering}. CRC Press.

\bibitem{hilborn2000chaos} Hilborn, R. C. (2000). \emph{Chaos and Nonlinear Dynamics: An Introduction for Scientists and Engineers}. Oxford University Press.


\bibitem{Braverman2013} E. Braverman, A. Rodkina (2013). \emph{Difference equations of Ricker and logistic types under bounded stochastic perturbations with positive mean}, Computers and Mathematics with Applications, Volume 66, Issue 11, 2013, Pages 2281-2294.

\bibitem{Aktar2023} Md Aktar, Ul Karim, Vikram Aithal, Amiya Ranjan Bhowmick, (2023). \emph{Random variation in model parameters: A comprehensive review of stochastic logistic growth equation}, Ecological Modelling, Volume 484, 2023, 110475.


\bibitem{Nasell2003} Ingemar Nåsell,(2003). \emph{Moment closure and the stochastic logistic model}. Theoretical Population Biology, Volume 63, Issue 2, 2003, Pages 159-168,

\bibitem{Pielou1975} E. C. Pielou (1975). \emph{Ecological Dirersity},  Wiley, New York, 1975.

\bibitem{Engen1978} S. Engen, (1978). \emph{ Stochustic Abundunce Models}, Chapman and Hall, London, 1978.


\bibitem{Schreiber2021} Schreiber, Sebastian J. and Huang, Shuo and Jiang, Jifa and Wang, Hao (2021). \emph{Extinction and Quasi-Stationarity for Discrete-Time, Endemic SIS and SIR Models}, SIAM Journal on Applied Mathematics, 81(5) 2195-2217.


\end{thebibliography}
\end{document}